# Go4 v2 Analysis Framework

J. Adamczewski, M. Al-Turany, D. Bertini, H.G. Essel, S. Linev
*GSI, 64291 Darmstadt, Germany*

Go4 developed at GSI is an analysis framework with a general purpose non blocking GUI. Go4 is based on ROOT. The GUI is implemented in Qt using GSI's QtROOT interface. Analysis and GUI run in separate tasks communicating through asynchronous object channels. A Go4 analysis may use any ROOT features. It can be organized in steps, which can be controlled from the GUI according to the user specifications. Each step is composed of event objects, the event processing, and event IO. Go4 composite event classes allow the construction of arbitrary complex events hierarchically composed of objects. The IO of the composite event objects to and from ROOT trees/branches is provided without explicit programming. The Go4 tree viewer can browse arbitrary hierarchy levels of composite events. The GUI provides hooks to attach user written GUIs. These GUIs have access to all objects of the analysis, i.e. events for asynchronous event display. Using the Qt designer the development of such GUIs is very efficient. The Go4 fit package (API and GUI) is a powerful and extendable tool to model and fit experimental data. The Go4 framework is especially useful for on-line monitoring. The HADES experiment at GSI integrated the existing ROOT based analysis into Go4 and is using it on-line with dedicated GUIs.

## 1. OVERVIEW

The GSI on-line off-line object oriented analysis framework Go4 is based on ROOT [1]. Additionally, the Go4 GUI uses heavily the Qt graphics library [2], which interoperates with ROOT using the QtRoot interface developed at GSI [3]. Main features are:

- A framework for many experiments (AP & NP)
- The analysis is written by the user (unlimited ROOT)
- A GUI controls and steers the analysis
- The GUI is never blocked by analysis
- The analysis may run permanently (e.g. on-line), is never blocked by GUI
- An analysis may update graphics asynchronously
- The analysis may run without change in batch mode

Different modes of operation exist for the Go4, depending on the degree of adjustment of the user analysis to the framework handles.

## 2. GO4 GUI AS STANDALONE TOOL

The multiple document interface GUI (MDI) of the Go4 can be used without any user analysis for interactive analysis of ROOT files. Beside the standard tools of ROOT, the Go4 GUI delivers a series of extra widgets that can be used to interact with Go4 objects (i.e. parameters, conditions, etc…).

### 2.1. Go4 Disk Browser and Tree Viewer

The disk browser and the Tree viewer of the Go4 can handle any ROOT based files and trees; moreover the tree viewer can also handle trees containing a Go4 composite event.

### 2.2. Go4 Histogram Client

The histogram client of the Go4 is used to connect to the existing GSI packages that deliver a histogram server, like GOOSY, MBS, and LeA [4]. The histograms (having their own format on the different servers) are converted to ROOT histograms on the client side, thus they can be modified, and saved into ROOT files afterward. Additionally, this client can connect to the optional histogram server in the Go4 analysis task (see below), so any standalone Go4 GUI may fetch histograms from any running Go4 analysis via this histogram client.

### 2.3. Go4 Fitter

The Go4 fitter package is an independent add-on to ROOT. It features a modular extensible design capable of using any kind of minimizer, fit function, and model, inside and outside ROOT [5]. The Go4 GUI offers a Fitter window to perform fits on any 1- and 2- dimensional histograms in memory or in a ROOT file.

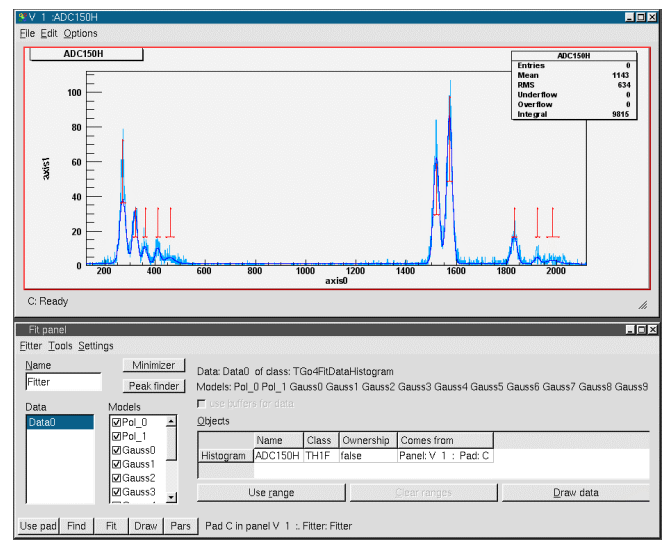

Figure 1: Fit panel with example fit

Automatic peak finders allow a fast setup of fit models and initial parameters for a given histogram. All peaks in figure 1 have been set up by the peak finder. Fit and results can be controlled both in an intuitive wizard mode, and in an expert mode with full access to the entire functionality of the Go4 Fit API.

## 3. GUI TASK AND ANALYSIS TASK

### 3.1. Inter-task Communication

To control an on-line or off-line analysis without blocking from the GUI, analysis and GUI run in separate tasks connected via sockets and communication threads [6]. Figure 2 shows a schematic view of the tasks with their functional components. A complete object transfer between both tasks is provided by means of the ROOT streamer mechanism. Exchange of command, data, and status objects allow control of the analysis task from the GUI. Any ROOT object from the analysis side can be required from the GUI side. Both, the standard Go4 remote object browser, and a user written GUI may request and receive analysis objects. Such objects also can be sent asynchronously by the analysis.

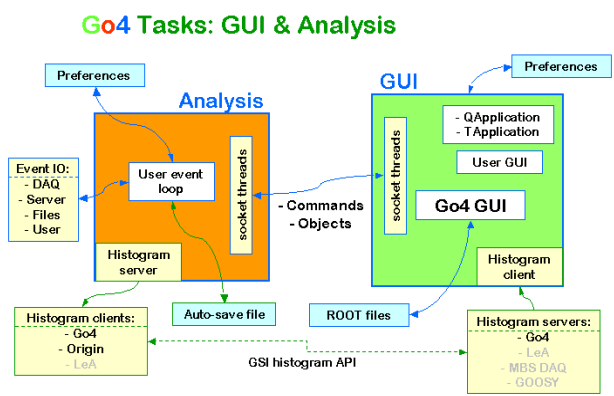

Figure 2: Analysis and GUI tasks

### 3.2. The GUI Registry

The dispatching of the incoming objects on the GUI side is handled by a single GUI Registry instance. Commands that request an object from the analysis register here which action shall happen when the object arrives, e.g. drawing a histogram in the specific panel, or passing a condition to the editor window. The registered action will be executed on arrival synchronized with other GUI activities. Moreover, in the same manner it is possible to generally redirect an object by its name to the optional user GUI. This makes it possible to handle user written objects coming from analysis without request, e.g. online event information that are only sent when an "interesting" event occurs.

### 3.3. Histogram Server

Besides the one-to-one connection between analysis and GUI, histograms of one analysis may be requested for visualisation by many GUIs. This is achieved by a histogram server/client mechanism using a standard GSI histogram API with tcp sockets. This interface enables histogram exchange between different environments, such as Go4, Microcal ORIGIN, the MBS DAQ, and the previous GSI analysis frameworks LeA and GOOSY.

The histogram server implemented in the analysis task (Figure 2) allows to export histograms to all of these. Vice versa, any Go4 GUI can import histograms from these by the histogram client pad. By this mechanism any number of standalone Go4 GUIs may visualize histograms from one or several Go4 analysis tasks independently.

### 3.4. Object Server

Since the GSI histogram API is dedicated for histograms only, it can not transport other ROOT objects, e.g. conditions, parameters, or user defined ones. To access these from a running analysis, Go4 offers an additional object server with a ROOT streamer and socket mechanism similar to the inter-task connection between analysis and controlling GUI. A user written ROOT application may apply the corresponding object client class to request any analysis object from this server.

## 4. ANALYSIS FRAMEWORK

The Go4 analysis framework can run both in batch mode or in GUI controlled multi-threaded mode. Any kind of user analysis shall be treatable by the framework. This is achieved by an object oriented approach.

### 4.1. Base Classes

User analysis and event class implementations inherit from a set of Go4 base classes. These define interfaces for the framework to handle any analysis in a common manner. Base classes exist for event structures, event processors (algorithms), event IO (source and store), and the analysis frame itself. Classes for standard GSI event sources and several types of events are provided. These events are stored/retrieved in ROOT trees. The implementation of other event sources is supported by templates.

### 4.2. Analysis Steps

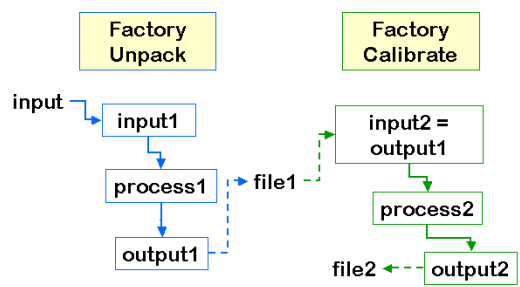

Figure 3: Analysis steps data flow

The Go4 analysis is organized in steps. Each step has an event source, an input event structure, an event processor, an output event structure, and an event store. Figure 3 shows two steps "Unpack" and "Calibrate". These are implemented by the user and are created on initialization time by means of a factory instance for each step. The steps can run subsequently, each working on the output event of the previous one. Moreover, each step can work independently, getting input from its own event source, and writing output into its own event store. The execution of the steps and the event IO can be controlled from the GUI.

### 4.3. The Go4 Composite Event.

The composite event classes can be used to build complex hierarchically structured event objects reflecting the natural experimental setup. Complex events are built out of simple data objects and/or composition of data objects uniformly. Fast direct access of components is realized by indexing. Full or Partial IO is achieved by mapping the event object into a ROOT TTree. Composite events are stored using standard ROOT IO mechanisms without changes in TTree or TBranchElement. Event store/retrieval, fully or partially, is done in the base classes by recursive mechanisms without the need of extra user written code. The Go4 tree browser resolves the tree hierarchy up to unlimited levels.

## 5. GUI CONTROLLING THE ANALYSIS

### 5.1. Analysis Control

A Go4 analysis can run in batch mode, command line mode, or in the ROOT CINT interpreter. Or it can be started from the GUI. In this case the complete analysis setup can be controlled. The analysis step control window is formatted according to the steps defined by the user (Figure 4).

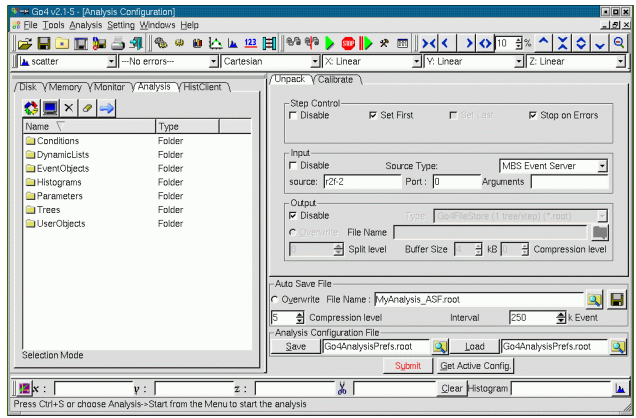

Figure 4: Analysis step control panel

### 5.2. Remote Object Browser

Any ROOT object, which is registered in the Go4 analysis, can be viewed and requested from the GUI via a remote object browser. Depending on the object type, it can be displayed in a Go4 view panel, or the corresponding editor window opens to modify its values, respectively.

### 5.3. Condition Editor

The Go4 analysis conditions are a set of classes dedicated to test values against given boundaries (windows or polygon shapes). These can be visualized and edited from the GUI (Figure 5). Conditions can be inverted or frozen to be always true or false, respectively. They increment test and true counters for statistics of the condition results.

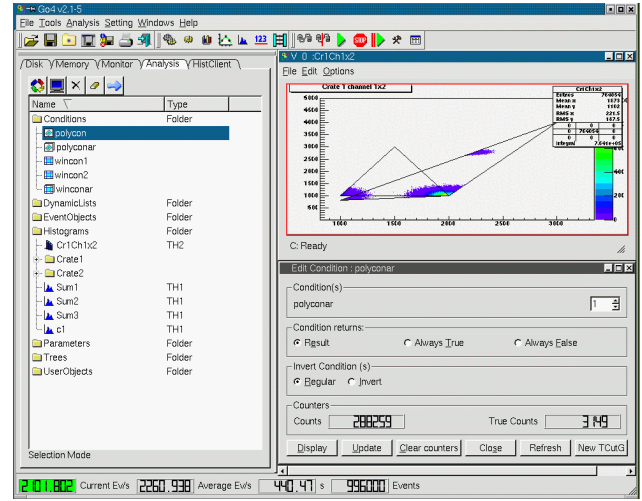

Figure 5: Condition editor

### 5.4. Parameter Editor

The TGo4Parameter base class offers a mechanism to edit user defined structures of values on the GUI and apply them in the analysis for any purpose. The user parameter subclass may contain any kind of supported basic types or arrays of them.

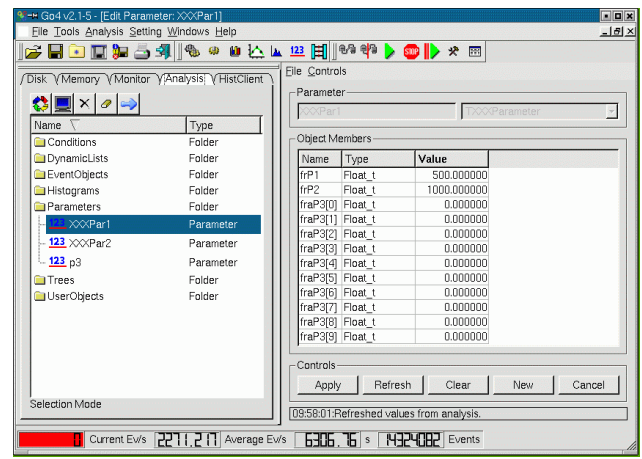

Figure 6: Parameter editor

These are evaluated automatically for display in the editor, using the ROOT class information, as seen in figure 6 on the right side.

### 5.5. Histogramming on the Fly

The Go4 dynamic list is a mechanism to define connections between histograms, conditions, and event data values on the fly without recompiling the user analysis. This is achieved by two alternative approaches: 1. using the ROOT TTree::Draw mechanism, when a tree exists in the analysis; 2. by direct pointer access to the event object members using the ROOT dictionary meta information. Both variants can be controlled from the GUI.

## 6. USER DEFINED GUI

Go4 offers hooks to attach a user defined GUI to the main Go4 window, with the possibility of full interaction with the analysis, i.e. sending commands, redirecting analysis objects to the user GUI, etc. The user GUI can be created and edited easily with Trolltech's Qt designer tool delivered with Qt [2].

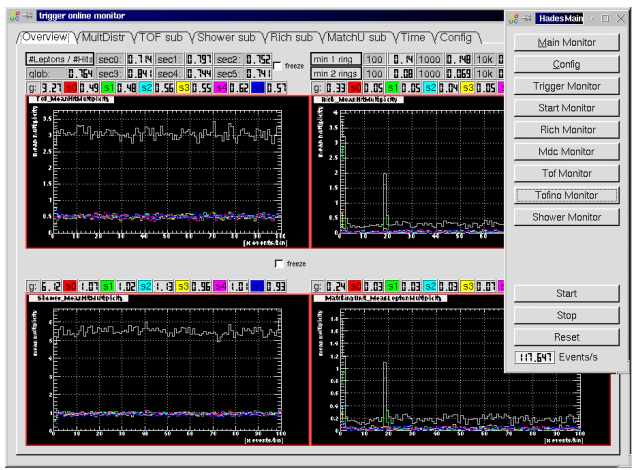

Figure 7: HADES online display user GUI

### 6.1. Example: Hades On-line Monitoring

The Hades collaboration [7] used the Go4 framework to implement on-line monitor of the different parts of the Hades detector. The existing ROOT based analysis have been embedded into the Go4 framework. The analysis is controlled by specific GUIs implemented in Qt using the Go4 hooks.

Figure 7 shows a screenshot. The information in the graphics window is updated from the analysis continuously.

## 7. CONCLUSIONS

Go4 is a versatile framework that has already proved to fulfill most of the requirements for the GSI medium sized experiments. It is still under continuous improvement. The standalone Go4 GUI can be used as an interactive analysis tool for all ROOT based data.

Any existing ROOT analysis can be adopted to run inside the Go4 analysis event loop, with the benefit of non-blocking GUI control from a remote task. A new analysis can be implemented within the Go4 analysis steps logic, and may run both in batch or GUI mode. A user defined GUI is possible with full access to all analysis objects.

Go4 v2 is now available for download from the web site http://go4.gsi.de/. It is tested on Linux (Debian 2.2, Debian 3.0, RedHat 8.0, Suse 8.1) with several compilers (gcc 2.9, gcc 3.2, icc 7.0).